   \documentstyle[twocolumn,aps,eqsecnum]{revtex}
\tightenlines

\begin{document}

\draft
\title{Interaction-Free
Measurements, Atom Localisation and
Complementarity}
 
\author{ Anders Karlsson, Gunnar Bj\"ork and Erik Forsberg}

\address {Laboratory of Photonics and Microwave
Engineering, 
 Department of Electronics,\\ 
Royal Institute of Technology (KTH),
Electrum 229, 164 40 Kista, Sweden\\
electronic mail:andkar@ele.kth.se} 
 
\date{\today}

\maketitle

\begin{abstract}
We analyse interaction-free measurements  on classical and quantum objects.  We show the transition
from a classical interaction free measurement to a quantum  non-demolition measurement of atom
number, and discuss the  mechanism of the enforcement
of  complementarity in atom interferometric interaction-free measurements.
\end{abstract}


\narrowtext
 
\section{Introduction}
 \label{sec:introduction}

One of the conceptual foundations of
quantum mechanics states that one cannot obtain
 information about an object without the object
 being disturbed by the measurement process. Also the non-observance
of a result represents additional information and hence modifies the wavefunction, as 
described 1960 by Renninger, who used the notion of "negative result measurement" \cite{Renninger},
and later Dicke who analysed the change of an atomic wavefunction by the non-scattering of a photon
\cite{Dicke}. 
Elitzur and Vaidman (denoted EV in the following) \cite{ElitzurI}  pointed out the possibility 
of an "interaction-free"
measurement (abbreviated IFM in the following) in which the existence of 
an object is inferred without any photons
 "interacting" with the object. 

 Most treatments on IFM have considered 
classical objects, i.e.  reservoirs, thereby exluding any discussion about
backaction on the object. Here we will analyse IFM
on quantum objects, two level atoms and show that
an interaction free measurement on an atom in the dissipation free regime becomes a
quantum non-demolition measurement of atom number. We also show the enforcement of complementarity,
which follows {\em solely from   entanglement} \cite{Englert} using the IFM as  a delicate "which path"
("welcher weg") detector in an atom interferometric \cite{Mlunek} Young's double slit experiment
similar to the micromasers discussed in \cite{Walther}.

 Let us first discuss "which-path" detection in the scheme of  Elitzur-Vaidman (EV)
\cite{ElitzurI}, as applied to the coupled atom-photon Mach-Zender interferometers in Fig. \ref{double
interferometer}. In the EV scheme,  50 \% beamsplitters are used, and the interferometer is
balanced to give unity transmission into port $T$ when the atom is not in the interaction
region. A  maximum likelihood estimate ${\cal L}_m$ of detecting the atom path gives that
in half the
cases the atom will be in the lower atom-interferometer arm and
 the photon will hit detector $T$
with unity probability. In $1/4$ of the cases, the photon and atom will enter the same arm and
the photon will be absorbed. Let us denote this event by ${\cal A}$. If the object is in the upper arm,
but the photon chooses the upper photon interferometer arm, the second photon
beam-splitter will randomly scatter the photon to detector $R$ or $T$, each
event happens with probability 1/8. If
detector $R$ clicks we know the atom is in the interaction
arm, if detector $T$ clicks we cannot separate the event from
the case when the atom was in the non-interaction path.
 Hence, the likelihood estimate ${\cal L}_m = 1/2+{\cal A} + 1/8 =
0.875$ and ${\cal L}_m -{\cal A}= 5/8=0.625$. Note that ${\cal L}_m=1/2$ corresponds to no information
about the location of the object, one may equally well toss a coin.

 The 50 \% beamsplitter Mach-Zender is
not the optimum two-beamsplitter scheme in  terms of "which-path" detection.  For instance, varying
the second beamsplitter transmittivity, the likelihood optimally is ${\cal L}_m  = (5 + \sqrt{5})/8
\approx 0.904$, using  an (amplitude) transmittivity $t = [(1+ 5^{-1/2})/2]^{1/2} \approx 0.85$ and
${\cal L}_m -{\cal A} \approx 0.654$. Earlier analyses \cite{ElitzurI}, optimised the fraction of
interaction-free measurements, given that an object was in the path. We believe that optimising 
${\cal L}_m -{\cal A}$,
 the ability to detect an object without absorbtion, given no apriori information
about the location, is a good figure of merit for IFM.

Let us briefly discuss the quantum properties of the
 setup in Fig. \ref{double
interferometer}, sticking to the 50 \% photon, 50 \% atom beamsplitter case, the latter
constructed via one $\pi/2$- and one $\pi$-pulse for the atom. The
input is a single photon and a single ground-state two level atom. Let
$|g-\rangle,|e-\rangle$ denote  a ground or excited atom in the upper path, and $|-g\rangle$
the ground state atom in the lower path, i.e. the atom is never excited when taking the lower path.
 The joint
state vector after the beam-splitters is $( |1,0\rangle  + i
 |0,1\rangle)/\sqrt{2} \otimes (| g,-\rangle + i |-,g\rangle)/\sqrt{2} $, where
trivial, overall phase-factors are omitted. 
 The free evolution Hamiltonian reads
\begin{equation}
\hat{H}_0 = \hbar \omega [(\hat{a}^\dagger \hat{a} + {1 \over 2}) +  {1 \over 2}
(1 + \hat{\sigma}_z)  ] ,
\label{gunnar's free hamiltonian}
\end{equation}
\noindent
where  $\omega$ is the atom transition and light (angular) frequency, and $\hat \sigma_z= |e \rangle 
\langle e |-|g \rangle  \langle g  |$ is the atomic inversion.
The lower photon arm  and the upper atom arm in the interferometers
subsequently interact under the (rotating wave)
interaction Hamiltonian
\begin{equation}
\hat{H}_i = \hbar  \Omega_R (\hat{a}^\dagger \hat{\sigma}_- + 
\hat{a} \hat{\sigma}_+) .
\label{gunnar's interaction}
\end{equation}
Here  $  \hat \sigma_- =|g\rangle \langle e|$, and $ \hat \sigma_+ =|e\rangle \langle g|$ are the Pauli
spin-flip operators, $\hat a$ the photon annihilation operator,
and $\Omega_R$ the vacuum Rabi frequency.  Propagating through
the rest of the photon interferometer, generally renders a partially entangled
superposition of photon, ground- and excited atom states. However, if $\Omega_R \tau= \pi, 3
\pi,..$, the joint photon-atom state becomes perfectly entangled
  $| \psi_{at,det}\rangle =(-i|0,1,g,-\rangle +i
|1,0,-,g\rangle)/\sqrt{2}$, whereas if $\Omega_R \tau= 0, 2
\pi,..$, the joint
state becomes a product state 
$| \psi_{at,det}\rangle =|1,0\rangle \otimes (|g,-\rangle + i |-,g\rangle)/\sqrt{2}$.
Assuming perfect photodetection, when $\Omega_R \tau= \pi, 3 \pi,..$, the photon-atom
interferometer  makes a sharp (stroboscopic) quantum non-demolition measurement \cite{BraginskyII}
of the ground state atom number $\hat N_g =  |g-\rangle \langle g-|$, using an
absorbative coupling instead of a dispersive coupling usually employed. 
Note that the entanglement is an absolute necessity for any information extraction.
In the product state, there is no atom "which-path" information in the photon state.
Furthermore, note that when specifying the interaction Hamiltonian, the spatial overlap of the object
and the probe system by neccessity enters. Any non-zero readout eliminates all atom modes whose overlap
is zero, even though the measurement is not an explicit position measurement. This is why the IFM and
other quantum measurements also becomes a measurement of the "precense" of an object. Adding
losses (atom damping) to the systems, translates the result towards the classical
IFM.

The single Mach-Zender IFM scheme features a relatively large probability of
absorption.  Using a folded
Mach-Zender interferometer \cite{KwiatI}, or a cavity resonator
\cite{Yoshi,Haroche,KarlssonI} the likelihood may approach unity, i.e.
the "which-path" information can assymptotically be extracted without any photon
absorbtion--"interaction", or classical photon "precense" at the object. It could be noted that also the
folded Mach-Zender or the resonator can be further optimised \cite{Bjork}. Let us
 discuss one such scheme, the interaction free measurement in a symmetric 
resonator, replacing the photon interferometer by a cavity detector system as in  Fig.
\ref{Atomwg}.  
 
In the cavity IFM, a
symmetric cavity is put on resonance to transmit all photons when empty. With an object present, the cavity
interference is suppressed and the photons are (mostly) reflected (hence not entering the cavity),
indicating the presence of the object.
 For an opaque object, standard resonator formulas gives ${\cal L}_m=1$ and ${\cal
A}=1-R$, where $R$ is the resonator (power) reflectance. Thus ${\cal L}_m-{\cal A}=R \rightarrow 1$
is asymptotically feasible.  Furthermore, it is easily shown that in the cavity IFM, the average
dissipated energy $\hbar \omega {\cal A}$ times the observation time $T_{meas}$ fulfills $\hbar
\omega {\cal A} \times T_{meas} \geq \hbar/2$, Bohr's energy-time uncertainty relation. We believe this is a
lower limit for any IFM scheme.

 To generate a single photon input we use an auxillary high Q cavity
with a single photon inside that is unidirectionally coupled to the IFM cavity,
i.e. a cascaded quantum system \cite{Gardiner,Carmichael}. The 
 Hamiltonian becomes

\begin{eqnarray} \hat H_0 = {\hbar \omega } [ {1\over 2}( 1+\sigma_z )
   &+&(\hat a^\dagger \hat a + {1 \over 2}) 
 +( \hat b^\dagger \hat b + {1 \over 2})]\nonumber \\+ \hbar  \Omega_R (\hat{a}^\dagger \hat{\sigma}_- + 
\hat{a} \hat{\sigma}_+) &+& i \hbar{1\over 2} \sqrt{\gamma_a \gamma_b }(\hat a^\dagger \hat
b-\hat b^\dagger \hat a) ,\end{eqnarray}

\noindent
where $\hat b$ is the photon annihilation operator in the auxillary cavity, and the last term in the
Hamiltonian describes the direct cavity coupling. The
irreversible photon transfer to the detectors is described by the Lindblad operators $\hat C_R =
(\sqrt{\gamma_{a}}  \hat a + \sqrt{\gamma_b} \hat b) \hat d_R^\dagger,$ and $ \hat C_T = \sqrt{\gamma_{a}} 
\hat a \hat d_T^\dagger$, where $\gamma_b$ is the decay rate from the auxillary cavity, $\gamma_{a}$ is the
single sided decay rate from the IFM cavity, and $\hat d_R$ and $\hat d_T$ represent the detector
states. 

 To quantify the information extraction and the backaction,
we   use the general interferometric quantities of Englert \cite{Englert} of {\em
distinguishability} ${\cal D}$, {\em maximum likelihood} ${\cal L}_{opt}$, and {\em fringe
visibility} ${\cal V}$.  Let $\hat \rho =\hat \rho_u+\hat
\rho_l $ be the density matrix of the coupled atom-photon system, where
$ \hat \rho_u = \langle g-| \hat \rho |g-\rangle+ \langle e-| \hat \rho
|e-\rangle  $ and $ \hat \rho_l =   \langle -g|\hat \rho |-g\rangle$, are the (unnormalised) density
matrices for the atom taking the upper or lower path, respectively. 
The distinguishability $\cal D$
can be defined \cite{Bjork}, slightly generalised from  \cite{Englert}, as
\begin{equation} {\cal D} \equiv  Tr_{det} \{ |(\hat \rho_u-\hat \rho_l)|\}, 
\end{equation}
\noindent
where $ |\hat A|\equiv \sqrt{\hat A^\dagger \hat A}$ denotes the absolute value of the operator
$\hat A$ and the trace is taken over the detector states.
From this, the optimum maximum likelihood ${\cal L}_{opt}$, {\em i.e. making the optimum use of the
information in the probe states} is given from \cite{Englert},
 $ {\cal L}_{opt} = {(1+{\cal D})/ 2} $. Note, that what is measured  in an actual experiment is the Hilbert space
distance in a chosen detector basis $ \{ |\psi_i \rangle \}$ 

  \begin{equation} {\cal D}_m \equiv \sum_i |\langle \psi_i|(\hat \rho_u-\hat \rho_l)|\psi_i\rangle| \leq
{\cal D} \label{D_M}, \end{equation}
giving the corresponding likelihood estimate ${\cal L}_m\equiv (1+{\cal D}_m)/2 \leq {\cal L}_{opt}$
\cite{Englert,Bjork}. 
 To quantify the backaction on the atom state we use the {\em atomic
fringe visibility} $\cal V $ given from the reduced
density operator $\hat \rho_{at}\equiv {\rm Tr_{det}} \{ \hat \rho  \}$
as ${\cal V} \equiv 2 |\langle g,-|{\rm Tr_{det}}\{ \hat \rho  \} |-,g \rangle|=2 |\hat \rho_{at,-g}| $.
The quantities ${\cal D} $ and ${\cal V}$ satisfies the duality relation \cite{Englert},

\begin{equation} {\cal D}^2 +{\cal V}^2 \leq 1, \label{Dual} \end{equation}
\noindent
 with equality only for a detector initially in
a pure state, and the entangling interaction  being unitary. This 
relation is a fundamental statement of complementarity. 

 In Fig. \ref{VisDplot}  the fringe visibility $  {\cal
V}$, the maximum likelihood ${\cal L}_m$  and the square sum ${\cal V}^2+{\cal
D}_m ^2$ are plotted as a function of the normalised
interaction time $\Omega_R \tau$. The parameters are $\gamma_a/\Omega_R=0.4,
\gamma_b/\Omega_R=0.04$, where $\gamma_a>> \gamma_b$ is needed in order for the photon to fully enter the
IFM cavity, and $\Omega_R>2\gamma_a$ is needed to have a reflection when the atom is inside. As seen,
the cavity IFM localises the atom with a high efficiency. Note that, classically the photon
(asymptotically with increasing reflectivity)  does not enter the cavity, yet a measurement is
performed. The mechanism is that the impinging photon sets up the dressed
 atom-photon states $|\pm\rangle =1/\sqrt{2}[|g,1\rangle \pm |e,0\rangle $ with a large
reflection for the incident light at the cold cavity resonance. In this case, the
{\em single pass phase shift} (energy dressing) is very small, on the order of $\Delta \phi \propto
(1-R)$, compared to $\pi$ for the atom-photon interferometer. The reason why ${\cal V}^2+{\cal D}_m^2
<1$ stems from the choice of detection basis and the continous non-unitary observation by the detectors. 
For a discrete measurement, by a proper unitary interaction before the detection, one may reach ${\cal D}_m={\cal D}$
\cite{Bjork}. For the continous measurement case, it remains a bit unclear whether that is feasible. However, as ${\cal
D}_m \rightarrow 1$ when the reflectivity approaches unity, we believe one should be able to reach
also ${\cal D}_mÊ\rightarrow {\cal D}$ in this case.  For short interaction times $t <<
1/\Omega_R$, ${\cal D}_m \propto t^2$ and the absorbtion ${\cal A}\propto t^5$ (if losses are
inserted). This ability to extract information without absorbtion in a (repeated) weak measurement
is at the heart of the high efficiency IFM. It does not, however, imply the absence of backaction
(decrease in fringe visibility), as is evidenced from the duality relation Eq. \ref{Dual}. Note
that the visibility decreases according to the potentially available information ${\cal D}$, not
with that which actually is extracted ${\cal D}_m\leq {\cal D}$.

Let us discuss the conditioned dynamics of the succesfull IFM,
 i.e. as given from a readout in the $d_R$ detector. The atom
density matrix $\hat \rho_{at,cond}$  contingent
 on the readout of a photon in the $d_R$ detector
 is given from $ \hat \rho_{at,cond}=  \langle g,d_R| \hat \rho | g,d_R \rangle / | \langle
g,d_R| \hat \rho | g,d_R \rangle | $. In Fig. \ref{Condmeas}, is shown the conditioned evolution
of the reduced atomic density matrix initially prepared in the even superposition state 
$|\psi_{at,e} \rangle =1/{\sqrt 2}[|g-\rangle+|-g\rangle] $. The numerical parameters are
$\gamma_a/\Omega_R=0.4, \gamma_b/\Omega_R=0.04 $. Two cases are illustrated $\gamma_c=0$
corresponding to no spontaneous emission, and $\gamma_c/\Omega_R=1.2$ corresponding to  strong
atom damping, i.e. through the damping operator $\hat C_c  =
 \sqrt{\gamma_{c}} \hat \sigma_-$. Here there is still some probability of the photon first entering
the cavity, and then exiting to detector $d_R$. However, increasing $\gamma_c$ (and $\Omega_R$), 
the probability of the photon being in the cavity and not being absorbed decreases, but
the conditioned evolution remains similar.

Let us finally discuss the enforcement of complementarity in IFM on quantum objects. The cavity IFM
scheme is very similar to the micromaser "which path" detector/quantum eraser  \cite{Walther,Scully},
 where  "which path" information is obtained
without scattering or introducing large uncontrolled phase factors into the interfering
beams. The  post IFM atom density matrix in the ideal case of zero visibility can be written as a mixture of the
complementary interference patterns $ \hat \rho_{at,out}
=1/2 |\psi_{at,e} \rangle \langle \psi_{at,e} |+ 1/2  |\psi_{at,o} \rangle \langle \psi_{at,o} |$,
where $|\psi_{at,e} \rangle =1/{\sqrt 2}[|g-\rangle+|-g\rangle] $ and 
$|\psi_{at,o} \rangle =1/{\sqrt 2}[|g- \rangle -
|-g\rangle] $. To extract the pure states,  the detection basis must be shifted (before the detection) by a unitary
transformation, i.e. in practice by adding the reflected and transmitted beams on a 50 \% beamsplitter to give the new
detection basis $|d_R,d_T\rangle \rightarrow |1/\sqrt{2} (d_R+d_T),1/\sqrt{2} (d_R-d_T) \rangle $. In the {\em
conditioned dynamics} of selecting the outcomes from one of the two new detection basises, the interference is once again
retrieved. This is how the cavity IFM can be turned into a quantum eraser \cite{Scully}. General conditions for the
possibility to retrieve quantum interference in conditioned dynamics will be given elsewhere \cite{Bjork}.  
  The interesting difference compared to previous
quantum erasers is the  classical absence of interaction in the IFM between the probe photon and the atom.
In the micromaser "which path" detector  \cite{Walther,Scully}, the atom enters in the excited state
and exits in the ground state. Here the atom essentially remain in the ground state throughout
the interaction.  The coupled atom-photon interferometers also displays these features \cite{Bjork}, i.e. the
upper path atom wavefunction makes a deterministic $\pi$ phase shift if and only if the photon
takes the same path. We like to stress that {\em no net energy and no net momentum is imparted
on the atoms in the IFM schemes}. $|\psi_{at,e} \rangle $ and $|\psi_{at,o} \rangle $ have essentially
the same mean momenta (=zero) and roughly the same variance.
Complementary is enforced, not refering to the position-momentum uncertainty relation, but is due to the
entanglement in the atom-photon state established by the unitary interaction on the initial atom
superposition state, in accordance with the view of Scully et. al. \cite{Walther}.

In summary, interaction-free measurements are interaction-free in that the average absorbtion can be much
less than one energy quanta \cite{Gabor}. For quantum objects we may
always (in principle) observe the modification of the wavefunction also for the succesfull IFMs. Yet, the more
opaque and classical the object gets, i.e. the stronger the coupling to a reservoir becomes, the smaller
is the detectable trace of interaction, in the limit of a completely opaque classical object to the point where
there cannot have been an interaction for the succesfull IFM \cite{ElitzurI,KwiatI}. The mechanism of the enforcement of
complementarity is subtle, but follows from the (physical) process of the entangling unitary interaction.

\section{Acknowledgements}
The authors would like to thank Edgard Goobar, Tedros Tsegaye and
G\"oran Lindblad  at KTH, Yoshihisa Yamamoto at Stanford
University, Paul G. Kwiat at Los Alamos National Laboratory, and Harald Weinfurther and Anton
Zeilinger at Innsbruck University for useful discussions and for sending preprints.   This work was supported by NFR--the
Swedish Natural Science Research Council and TFR--the Swedish Technical Science Research Council.

${ } \quad$
\begin{figure}[b]
\caption{Coupled atom-photon interferometer using 50 \% atom- and 50 \% photon beamsplitters. The box
$ H_i$ denotes the interaction region. } \label{double interferometer}
\end{figure}

${ } \quad$
  \begin{figure}[b] 
\caption{Atom interferomenter with a cavity IFM detector in one of the atom arms.
}  \label{Atomwg} \end{figure}

 \begin{figure}[b] 
 

\caption{Fringe visibility ${\cal V}$,  maximum likelihood ${\cal L}_m$ and (non-optimum)
duality ${\cal D}_m^2+{\cal V}^2$ plotted as a function of normalised time.
}  \label{VisDplot} \end{figure}

\begin{figure}[b] 
 

\caption{Time evolution of the conditioned atomic density operator elements $\hat \rho_{at,gg}$ (
probability of atom in IFM path) and $|\hat \rho_{at,g-}|={\cal V}/2$,  conditioned on the detection
in $d_R$ versus normalised time. }  \label{Condmeas} \end{figure}

\end{document}